\documentclass{llncs}
  \usepackage[pdftex]{graphicx}
\usepackage[cmex10]{amsmath}

\usepackage{multirow}
\usepackage[noadjust]{cite}

\begin{document}
%
\title{An Adaptive Primary User Emulation Attack Detection Mechanism for Cognitive Radio Networks}
\titlerunning{An Adaptive PUE Attack Detection Mechanism for CRNs}
\authorrunning{Q. Dong, Y. Chen, X. Li, K. Zeng}

\author{Qi Dong\inst{1} \and
		Yu Chen\inst{1} \and
		Xiaohua Li\inst{1} \and
		Kai Zeng\inst{2} \and
        Roger Zimmermann\inst{3}}
	\institute{Dept. of Electrical and Computer Engineering, Binghamton University, Binghamton, NY 13902, \email{\{qdong3, ychen, xli\}@binghamton.edu} \and
	Volgenau School of Engineering, George Mason University, Fairfax, VA 22030, \email{kzeng2@gmu.edu} \and
     School of Computing, National University of Singapore, Singapore 117417, \email{rogerz@comp.nus.edu.sg} }



\maketitle

\begin{abstract}

The proliferation of advanced information technologies (IT), especially the wide spread of Internet of Things (IoTs) makes wireless spectrum a precious resource. Cognitive radio network (CRN) has been recognized as the key to achieve efficient utility of communication bands. Because of the great difficulty, high complexity and regulations in dynamic spectrum access (DSA), it is very challenging to protect CRNs from malicious attackers or selfish abusers. Primary user emulation (PUE) attacks is one type of easy-to-launch but hard-to-detect attacks in CRNs that malicious entities mimic PU signals in order to either occupy spectrum resource selfishly or conduct Denial of Service (DoS) attacks. Inspired by the physical features widely used as the fingerprint of variant electronic devices, an adaptive and realistic PUE attack detection technique is proposed in this paper. It leverages the PU transmission features that attackers are not able to mimic. In this work, the transmission power is selected as one of the hard-to-mimic features due to the intrinsic discrepancy between PUs and attackers, while considering constraints in real implementations. Our experimental results verified the effectiveness and correctness of the proposed mechanism.

\end{abstract}


%

\section{Introduction}
The rigid spectrum allocation scheme regulated by governmental agencies leads to great deficit on spectrum band resources. Static spectrum access technology results in lots of waste on wireless spectrum resources. The emergence of new intelligent spectrum allocation/re-allocation schemes, especially cognitive radio network (CRN), are studied elaborately in the last decade, due to the ever-increasing wireless applications. Cognitive radio (CR), or known as secondary user (SU) in CRN, is a technology that allows wireless devices (unlicensed users) access spectrum resources dynamically without introducing major interference to licensed primary users (PUs). Because of the great difficulty and high complexity in dynamic spectrum access (DSA), and many open issues on security deployment, CRN study still under development \cite{Zhang2017CNS}.

Spectrum sensing allows CRs acquire real-time spectrum occupation status such that interleaving communications shared by PUs and SUs become feasible. Basically, a well-designed CRN aims to serve for two purposes \cite{Adelantado:2013aa}: to maximize the usage of spare spectrum resource as well as to protect the incumbent primary system from secondary network interference. Due to the requirement to SUs that they shall not interfere the PU functionalities, SUs should adapt their behaviour in accordance to PU activities. Such requirement can be regarded as two separate parts: (1) monitoring PU activities, and (2) behaving properly.

In general, knowing PU activities is essentially critical for cognitive radios to share the spectrum resource with legitimate users. One of the effortless ways to acquire PU activity information is that PUs are able to notify SUs their spectrum usage status; or there exist a third party as an inquiry center that knows what PUs  will do in the near future. An alternative solution is to develop robust and efficient spectrum sensing technique to acquire knowledge on PU activities. Also, the spectrum sharing efficiency greatly depends on a secure CR operating environment. In addition, due to the opportunistic spectrum access (OSA) nature, CR systems encounter several CR-specified security problems.

Regarding spectrum sensing, one major challenge is to detect PU signals with high accuracy while maintain low false alarm rate. The false detection rate may become extraordinarily high when primary user emulation (PUE) attacks happen. A PUE attack is that malicious entities mimic PU signals in order to either occupy spectrum resource selfishly or conduct Denial of Service (DoS) attacks. PUE attacks can be easily implemented in CRNs. It introduces great overhead on cognitive radio communication and causes chaos in dynamic spectrum sensing \cite{dong2017anomaly,Dong:2017bb}. However, defense against the PUE attacks is nontrivial because traditional authentication and authorization (AA) methods are no longer applicable to CR systems. A more adaptive and practical PUE attack detection technique is highly desired.

Inspired by radiometric used to identify short range transceivers and the interpulse/intrapulse fingerprint in radar identification, we propose to detect PUE attacks in CRN environment leveraging the hard-to-mimic PU transmission features. As one type of hard-to-mimic feature, the PU transmission features are determined by the inherent physical characteristics of the device. Attackers are not able to generate such kind of features. A received signal strength (RSS)-based hypothesis detection mechanism is designed, which can detect attackers who attempt to fool the system by mimicking PUs' patterns.

In general, RSS-based approaches have been studied elaborately in many literatures for PUE attack defense. It is applied either as one direct rudimentary feature of PU \cite{Chen:2008aa}, or as the premise for PU localization \cite{Huang:2010aa, Das:2013aa, Marinho:2015aa}. These works can be challenged by either smart attackers or the practical constraints such as SUs are unaware of their geographical information. There are two major advantages that make our work more feasible and efficient in real-world applications than exiting solutions: (1) in general, our proposal allows mobility of nodes in the CRN and does not require prior geographical information of either PUs, SUs, or attackers; and (2) compared to machine learning or neural network based methods, our proposal does not need the training process.

The rest of this paper is organized as follows. Section \ref{sec:background} provides background knowledge that motivated this work. Section \ref{sec:model} describes a practical CRN model on which our detection mechanism is built. Section \ref{sec:perfect} discusses a PUE attack intuition under perfect propagation model assumptions. The proposed RSS-based PUE attack detection method is introduced in section \ref{sec:approach}. Section \ref{sec:experiment} presents a tentative trail based on real-world measurements. Section \ref{sec:evaluate} shows our numerical experimental results and comparison to other related schemes, and finally, Section \ref{sec:conclusions} concludes this paper.

\section{Background Knowledge and Related Work}
\label{sec:background}

According to Federal Communications Commission (FCC): \textit{``no modification to the incumbent signal should be required to accommodate opportunistic use of the spectrum by Secondary Users (SUs)''}  \cite{fcc2003}. Obviously FCC places constraints on PUs such that PUs are not obligated to notify CR users with their activity scheduling and intention, neither to provide AA services. Consequently, CR systems are expected to collect and process sufficient and highly accurate information of the spectrum environment without imposing overhead on incumbent users by adding new features, such as redundant symbolic pads or authentication protocols.

In CR systems, it is necessary to distinguish attacker signals from PU signals in spectrum sensing stage. PUE attacks will cause severe problems on the efficiency of spectrum utility. Since no obligation is imposed on PUs, it is natural to explore the features of different wireless transceivers. In general, there are two categories of transceiver features: the primary/strong radiometric/fingerprint, and the secondary/weak radiometric. The primary radiometric denotes the intrinsic characteristics or imperfections of wireless transceivers, that can be used to identify the  uniqueness of the hardware. Transient is one of the most discussed radiometric that can be used to identify short range transceivers. Transient is the part of the signal where the amplitude rises from background noise to full power. In literature, five transient features are used \cite{Rasmussen:2007aa}:

\begin{enumerate}
   \item The length of the transient, along the x-axis;
   \item The variance of the normalized amplitude of the transient;
   \item The number of peaks (periods) of the carrier signal in the transient;
   \item The first part of a discrete wavelet transform of the transient; and
   \item Difference between the normalized mean and the normalized maximum value of the transient.
\end{enumerate}

It is proved that transient features are useful fingerprints for wireless transceivers identification. They are not well studied in PU recognition in CRNs, however, due to the difficulties in detecting transient on the scale and scope of CRNs.

Another inspiration comes from radar identification, in which two kinds of fingerprint are usually discussed. One is interpulse fingerprint that considers factors including frequency, amplitude, pulse width, pulse repetition rate, etc. The other one is intrapulse fingerprint that pays attention to pulse waveform characteristics, such as unintentional modulation on pulse (UMOP) feature \cite{Langley:1993aa} and time domain waveform feature, including rise slope and fall time, falling angles, angle of pulse, and pulse point \cite{Kawalec:2004aa}. It looks intriguing, but requires accurate measurements on signals that is usually not available for CRs.

There are other ideas based on the imperfections of transceivers such as frequency offset error caused by different transmitter and receiver oscillators, or modulation errors caused by the imperfection of electric circuits \cite{Brik:2008aa}. Usually, those fingerprint extraction requires prior knowledge of modulation/mulplexing technology, and it is often very computational intensive.

The secondary/weak radiometric usually does not identify signals from a particular transceiver. Instead, it identifies signal characteristics that are not reproducible to attackers. A smart attacker is able to mimic some PU signal features such as spectrum bandwidth, activity pattern, and adaptively change transmission power. Many studies tried to extract features of communication channel of the wireless environment \cite{Chen:2008aa, Chen:2009aa, Huang:2010aa}, which is known as geometrical information of the PU transmitter, because PUs and attackers are unlikely be at the same place.

Two types of channel fingerprint detection approaches are well discussed. The first category is distance-based approaches \cite{Chen:2008aa}. A rudimentary approach is to use RSS-based location estimation techniques, which record the received energy level from the PU as the reference radiometric, and compare with the sensed spectrum signal strength for detection. A novel idea was proposed to deploy helper nodes around PUs, which are able to help verify PU signal based on helper node's authentic link signatures \cite{Liu:2010aa}. A smart attacker model was presented to prove that the first order feature of RSS is not adequate for PUE attacks detection, and then a RSS detection method using second order feature is proposed to confront the smart attackers~\cite{Chen:2008aa}. However, the assumption that all SUs and PUs' positions are prefixed and known is not applicable to many situations in CRNs. The second category is location-based approach \cite{Huang:2010aa}, which requires geographical information from at least part of network participators. In those proposals, peripherals such as GPS, helper nodes and prior knowledge of PU position, are necessary.

PUE attack detection happens in spectrum sensing stage. In 2010, FCC announced that they adopted condition a device's use of TV White Spaces on its consultation of a geolocation database to ensure the availability of the desired spectrum \cite{fccdatabase}. Several literatures have discussed the feasibility of constructing PU activity database and the details in design of prototypes \cite{Feng:2011aa, Murty2012, Pesko:2014aa, Yilmaz:2013aa}. The database will record, model and predict PU activities in order to regulate CR access and optimize spectrum use efficiency. These base stations are able to provide many critical PU information, such as geographical location, activity pattern, and modulation/mulplexing technology. Even further, a FCC Commission's Rule proposes that PUs such as Federal Primary Users are going to register in a database before accessing 3.5 GHz band \cite{fccDatabaseRegistry}.

On one hand, while such kind of database model can eliminate PUE attacks, they do violate the original FCC requirement \cite{fcc2003}. Database enabled spectrum sensing provides a new inspirations on against of PUE attack, but still remains problematic. As the general PU information is known to CRs with involvement of regular database, smart attackers can mimic PU signal features. In addition, the geographic information of PU is not available for moving base stations or radars. On the other hand, the PU registry approach has been deployed in very limited scale, which is only in federal PU environments \cite{fccDatabaseRegistry}.

As discussed above, a more adaptive and practical PUE attack detection technique is highly desired. Considering the limited prior knowledge of PUs and constraints on computing resources of CRs, it is natural to extend our vision on hard-to-mimic PU signal features for PUE attack detection. While the secondary radiometric can be easily reproduced by smart attackers, the actual transmission power is an exception. Although the attacker can smartly adapt their transmission power to disguise their locations, they are usually incapable of mimicking counterpart power as PUs. PUs are usually radars, TV stations, and cellular base stations, which signal strength is normally tens to thousands of times higher comparing to what PUE attackers can produce \cite{Paisana:2014aa}. For example, the strength of CRs signals is normally in scale of milliwatts \cite{Paisana:2014aa}. With cooperative spectrum sensing, and involvement of a fusion center (FC), the emitter transmission power based PUE detection is applicable without requiring any prior knowledge of PUs and CRs location information.

\section{Detection Model}
\label{sec:model}

In CR spectrum sensing study, the cooperative sensing method is preferable due to the well-known ``hidden PU problem''. This problem happens when a SU cannot sense an active PU either due to the PU signal is out of range or because the signal faded away in concurrent wireless fading channel. In cooperative spectrum sensing, CRs have to share their sensing results to obtain the most comprehensive knowledge of the desired spectrum environment. In centralized CRNs, a fusion center can collect and synthesize sensed spectrum information from all CRs, and make a joint decision on PU appearance. Our detection model is based on such deployment with the following assumptions.

\begin{itemize}
\item The PUs are either public infrastructures (i.e. TV stations) or federal facilities (i.e. weather radar system). They have powerful transmission capability to serve their own purposes.
\item The PUs are not required to be geographically fixed, such that PUs including moving radars or stations are considered.
\item Without loss of generality, assume CRs and the FC are randomly scattered in an circular area with radius of $r_{CRN}$. CRs are not equipped with localization peripherals, and they are unaware of the location of either themselves or the peers.
\item CRs are able to sense the radio environment and report processed spectrum features to the FC.
\item The FC can collect spectrum features from CRs and perform deliberate analysis. The FC has knowledge of general information of measured PUs, such as their occupied spectrum bands, their approximate propagation power, etc.
\end{itemize}

\begin{figure}[t]
	\centering
		\includegraphics[width=0.5\textwidth]{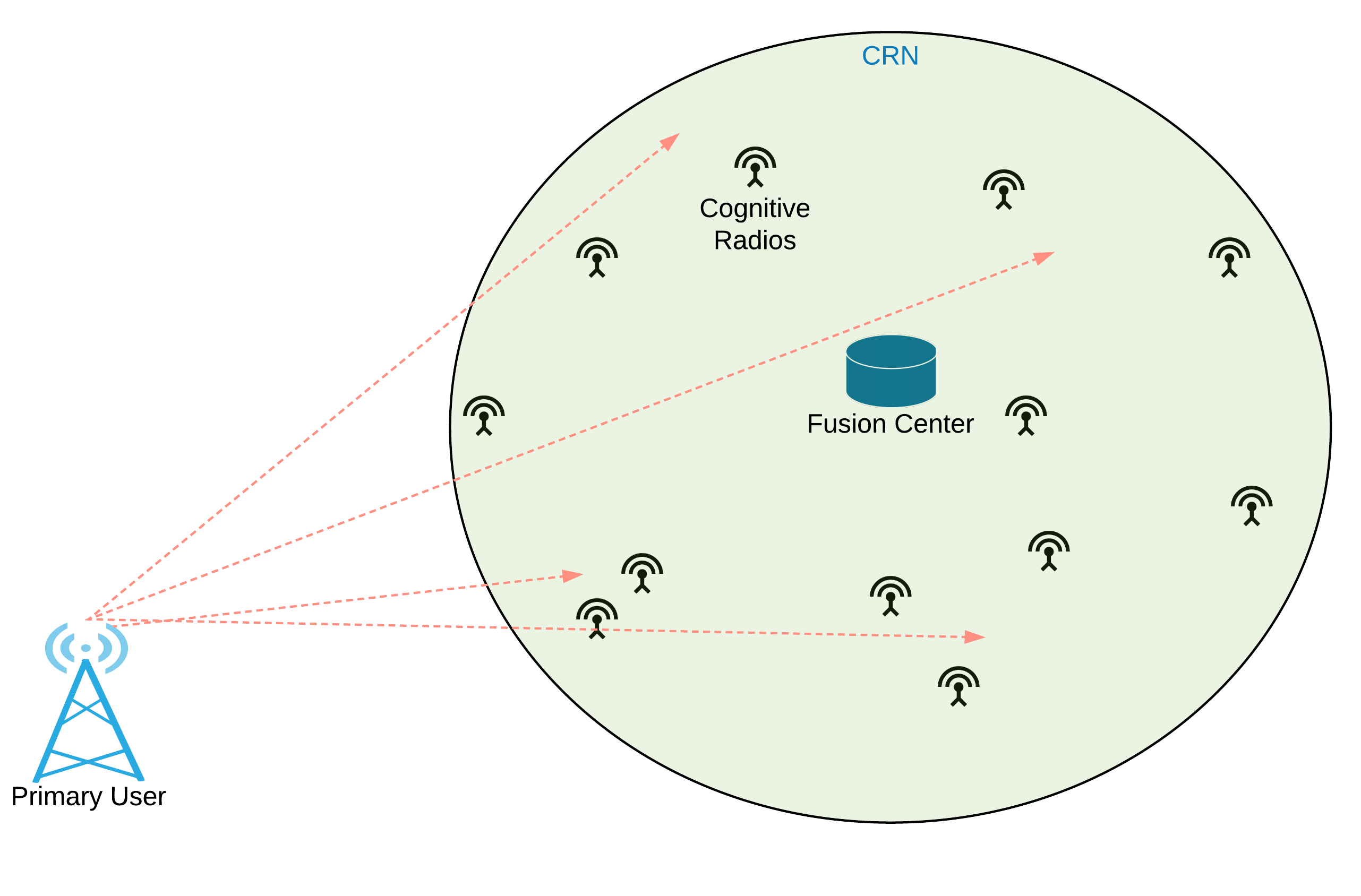}
	\caption{A centralized CRN sharing the spectrum with a PU.}
	\label{fig:model}
\end{figure}

Figure \ref{fig:model} shows a scenario of centralized CRN jointly share the spectrum resource with a PU. In order to be more consistent to real-world situation, in our detection model, the position of the PU and distances among each parties are unknown, and there is not localization peripherals, such as GPS, time of arrival (TOA) based equipment, is equipped by CRs because these peripherals are unaffordable in many applications. In consequence, this detection model poses a higher challenge on PUE attacks detection.

\section{PUE Attacks Detection under perfect propagation model}
\label{sec:perfect}

As discussed earlier, the attackers can hardly emit the magnitude of signal power as PUs do, so the propagation power becomes a useful hard-to-mimic secondary radiometric of transmitters. The challenge is, however, such a secondary radiometric feature remains unmeasurable in wireless environment. Usually, the receiver can measure the RSS, which is determined by many factors, such as transmission power, propagation environment, and transmission distance.

An ideal propagation model, Free-Space Path Loss (FSPL) model, assumes no obstructions between the transmitter and receiver, and the signal propagate along a line-of-sight (LOS) channel. This ideal propagation model inspires a reasonable intuition on PUE attacks detection. In this section, our new idea on PUE attacks detection is introduced with consideration of some restrictions in real world such as unknown PU and CRs locations, but we assume an ideal wireless propagation environment. The FSPL model is expressed as:

\begin{equation}
\label{eq:friis}
\frac{P_r}{P_t} =  \frac{G_l\cdot \lambda^2}{(4\pi d)^2}
\end{equation}

\noindent{where $P_r$ and $P_t$ are received signal power and transmitted signal power respectively; $\lambda$ is signal wave length; $d$ is the LOS distance between transmitter and receiver; $G_l$ is the product of the transmit and receive antenna field radiation patterns, and it is a constant if the pattern is known. Thus, the received to transmitted power ratio is proportional to the reciprocal of $d^2$ as:}

\begin{equation}
\label{eq:prop}
\frac{P_r}{P_t} \propto \frac{1}{d^2}
\end{equation}

\subsection{A Naive Detection Model}

In the ideal propagation model, given the RSS measurement and global information of PU propagation power, the transmitter-receiver distance is deducible, which gives us a hint on the relation between the uncloneable radio feature $P_t$ and the wireless channel feature $d$. In our PUE attacks detection model, a hypothesis test is adopted to decide the presence of the attacker.

\begin{description}
\item[] $\mathcal{H}_0: \quad$ the signal is from the PU
\item[] $\mathcal{H}_1: \quad$ the signal is from the attacker
\end{description}

The PU propagation power is usually in scale of hundreds or thousands of watts, defined as $P_{t,pu}$. In contrast, the attacker, usually comparable to CRs, has the propagation power of tens to hundreds of milliwatts, defined as $P_{t,attacker}$. Thus, the ratio of PU propagation power to attacker propagation power is computed as $R=P_{t,pu}/P_{t,attacker}$.

In a CRN with $N$ CRs, the transmitter-receiver distance $d_i$ ($i=1,0,\cdots,N$) can be easily computed given the propagation power $P_{t,pu}$ and individual CR received power $P_{r,i}$. If the signal is transmitted by the PU, the distance is computed as:

\begin{equation}
\label{eq:dPU}
d_{i}  =  M \cdot \sqrt{\frac{P_{t,pu}}{P_{r,i}}} = d_{i,pu}
\end{equation}

Here, $M$ is defined as a constant $M=\sqrt{\frac{G_l\cdot \lambda^2}{(4\pi)^2}}$. Similarly, if the signal is transmitted by the attacker, the distance is computed as:

\begin{eqnarray}
\label{eq:dATT}
d_{i} &=& M \cdot \sqrt{\frac{P_{t,pu}}{P_{r,i}}} =  M \cdot \sqrt{\frac{P_{t,attacker}}{P_{r,i}}}  \cdot \sqrt{R}\nonumber \\
	&=& d_{i,attacker} \cdot \sqrt{R}
\end{eqnarray}

\begin{figure}[t]
	\centering
		\includegraphics[width=0.5\textwidth]{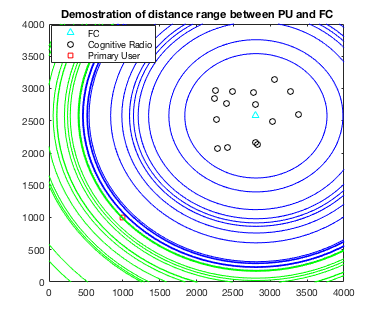}
	\caption{In the case of PU transmission, compute distance range between the transmitter and the FC. The radius of blue circles are the lower bounds of $d_{pu,fc}$, computed as $|d_{i}-d_{i,fc}|$; the radius of green circles are the upper bounds of $d_{pu,fc}$, computed as $d_{i}+d_{i,fc}$. The PU is supposed to locate between the lower and upper bounds.}
	\label{fig:pufcdistance}
\end{figure}

\begin{figure}[t]
	\centering
		\includegraphics[width=0.5\textwidth]{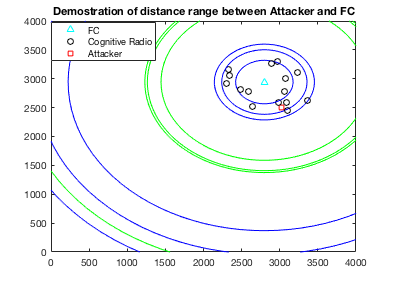}
	\caption{In the case of PUE attacks, compute distance range between the transmitter and the FC. The radius of blue circles are the lower bounds of $d_{attacker,fc}$, computed as $|d_{i}-d_{i,fc}|$; the radius of green circles are the upper bounds of $d_{attacker,fc}$, computed as $d_{i}+d_{i,fc}$. The figure shows no intersection between the lower and upper bounds.}
	\label{fig:puattdistance}
\end{figure}

Further, if the distance between individual CR and the FC $d_{i,fc}$ is also known, ideally, it is easy to infer to the distance between the PU and the FC in a range $d_{pu,fc} \in [max(|d_{i}-d_{i,fc}|) , min(d_{i}+d_{i,fc})]$. If the signal is transmitted by the PU, the computed $d_{pu,fc}$ does not belong to an empty set, as demonstrated in Fig. \ref{fig:pufcdistance}. If the signal is transmitted from the attacker, the distance is computed as $d_{i}=d_{i,attacker}\cdot \sqrt{R}$, according to Eq. \ref{eq:dATT}. Thus, the range set $d_{attacker,fc} \in [max(|d_{i,attacker}\cdot \sqrt{R}-d_{i,fc}|) , min(d_{i,attacker}\cdot \sqrt{R}+d_{i,fc})]$ is possibly empty as shown by Fig. \ref{fig:puattdistance}. The FC can apply the hypothesis test by:

\begin{itemize}
\item If $(d_{i}+d_{i,fc})\ge |d_{j}-d_{j,fc}|, \forall i,j=1,2,\cdots,N$,  the signal is from the PU ($\mathcal{H}_0$); or
\item If $(d_{i}+d_{i,fc})< |d_{j}-d_{j,fc}|, \exists i,j=1,2,\cdots,N$, the signal is from the attacker ($\mathcal{H}_1$).
\end{itemize}

Following the hypothesis test, the detection rate $P_d$ is calculated as:

\begin{eqnarray}
\label{eq:Prd}
P_d &=& 1-P_{fn} \nonumber \\
 	& \ge&  1-\Pr \{d_{i,attacker}-d_{j,attacker}\le \frac{d_{i,fc}+d_{j,fc}}{\sqrt{R}}, \nonumber \\
	&& \forall i,j =1,2,\cdots,N\} \nonumber \\
 	&\ge& 1-(1-(\frac{r_{CRN}-\frac{max(d_{i,fc}+d_{j,fc})}{\sqrt{R}}}{r_{CRN}})^2)^N,
\end{eqnarray}

\noindent{where $P_{fn}=\Pr(\mathcal{H}_0|\mathcal{H}_1)$ is the false negative probability. In Eq. \ref{eq:Prd}, the first inequality originates from the expansion of the inequality to the abosulote value of $|d_{i}-d_{i,fc}|$. The second inequality can be explained that the greatest false negative probability happens (suppose $\frac{max(d_{i,fc}+d_{j,fc})}{\sqrt{R}}\le r_{CRN}$) when the attacker is located in the center of CRN, and all CRs are located in the ring-shape area centered at the attacker with inner radius of $r_{CRN}-\frac{max(d_{i,fc}+d_{j,fc})}{\sqrt{R}}$ and outer radius of $r_{CRN}$. The false positive probability $P_{fp}=\Pr(\mathcal{H}_1|\mathcal{H}_0)$ is zero under such hypothesis test condition.

\subsection{Evaluation of hypothesis test by Monte Carlo method}

A Monte Carlo method is applied to calculate the detection accuracy in a scenario where CRs and attackers are randomly distributed in an circular area, which is centered at the FC with radius $r_{CRN}$. Figure \ref{fig:monte} shows the result. The detection accuracy $P_d$ increases dramatically as the number of CRs increases. And $P_d$ is approaching one when there are more than four CRs in the testing scenario.

\begin{figure}[t]
	\centering
		\includegraphics[width=0.6\textwidth]{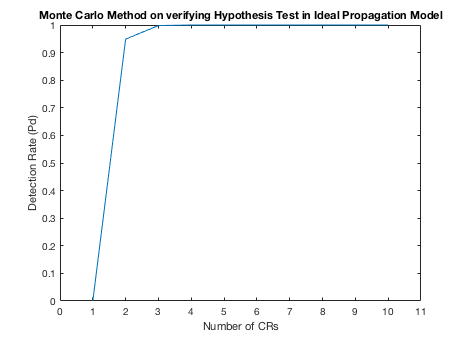}
	\caption{The detection accuracy of the hypothesis test computed by Monte Carlo method with different number of CRs in the CRN. For each different number of CRs, repeat the hypothesis test for 100000 times.}
	\label{fig:monte}
\end{figure}

\section{A RSS based PUE attack detection approach}
\label{sec:approach}

The above hypothesis test is discussed under ideal propagation model, which provides a reasonable intuition on PUE attack detection, with the given propagation power features of PUs and attackers. But, in reality, the RSS based distance measurement method is not well applicable for several reasons. First of all, the FSPL propagation model cannot faithfully describe the actual propagation environment. Secondly, signal propagation patterns are variant in different environments. Also, RSS can be vary by a large magnitude over short distances.

Therefore, we choose the single transmitter log-normal shadowing fading propagation model to describe the relationship among transmitted power $P_t$, received signal power $P_r$, and distance $d$ between transmitter and receiver.

\begin{equation}
\label{eq:lognormal}
P_r(dBm) = P_t(dBm) + K (dB) - 10 \gamma log_{10} \frac{d}{d_0} + G
\end{equation}

\noindent{where $P_t$ and $P_r$ are measured in $dBm$. $K$ is the path loss variable at the reference distance $d_0$, which depends on the antenna characteristics and propagation environment. $\gamma$ is the empirical path loss exponent, which is learned to have different values in different environment \cite{wirelessCommunication}. Table \ref{tab:gamma} presents some $\gamma$ values measured by empirical studies. $G$ is a normal random variable with zero mean and standard deviation $\sigma$. Most empirical studies for outdoor channels measure the standard deviation $\sigma \in (5,12)$ in macrocells and $\sigma \in (4,13)$ in microcells \cite{wirelessCommunication}.}

\begin{table}[ht]\centering
\caption{Empirical Path Loss Exponents $\gamma$}
\label{tab:gamma}
    \begin{tabular}{|p{30mm}|p{30mm}|}
    \hline
         Environment & $\gamma$ range \\
	\hline
        	Urban macrocells &  $3.7 - 6.5$  \\
        	\hline
         Urban microcells   &   $2.7 - 3.5$	\\
         \hline
	Factory   &   $1.6 - 3.3$	\\
         \hline
    \end{tabular}
\end{table}

Over the years of development, a number of propagation models have been developed in different wireless environments, such as Hata model, COST231 model, piecewise linear model, etc. \cite{wirelessCommunication}. In some literatures, a statistical model is used to obtain maximum likelihood of the propagation model parameters with great fitness \cite{Roos:2002aa}. In our work, we assume the model parameters with some errors, are accessible either from historically empirical study, or statistical estimation. Thus, the path loss propagation model, inferred from Eq. \ref{eq:lognormal}, can be written as Eq. \ref{eq:model}, where $C$ is a constant determined by reference propagation path loss, and $\Gamma$ is the empirical path loss exponent.

\begin{equation}
\label{eq:model}
L=P_t - P_r =  C+\Gamma \cdot log_{10} d+ G
\end{equation}

Because $G$ is a normal random variable, the optimal estimator of $log_{10} d$ is obtained by averaging the propagation loss $L$. Thus, we smooth the RSS by using a local averaging method from neighboring CR groups. Then, we apply our hypothesis test to detect PUE attacks.

\subsection{CRs Grouping}
A RSS smoothing method that divides secondary network into circular areas has been studied \cite{Chen:2008aa}. One major restriction of this method lies in the requirement that all CR positions are known globally and CRs remain geographically static. In our work, as discussed in Section \ref{sec:model}, a dynamic CRN is assumed where CRs can be either static or mobile, and the CRs are assumed unaware of their positions. In order to estimate distance to the PU in a small area, a CR grouping technique is applied, which assumes the distances between the PU and CRs in a group can be uniformly treated as $d_{i,pu}$, where $i$ represents the $i$-th CR as the group leader.

\begin{figure}[t]
	\centering
		\includegraphics[width=0.7\textwidth]{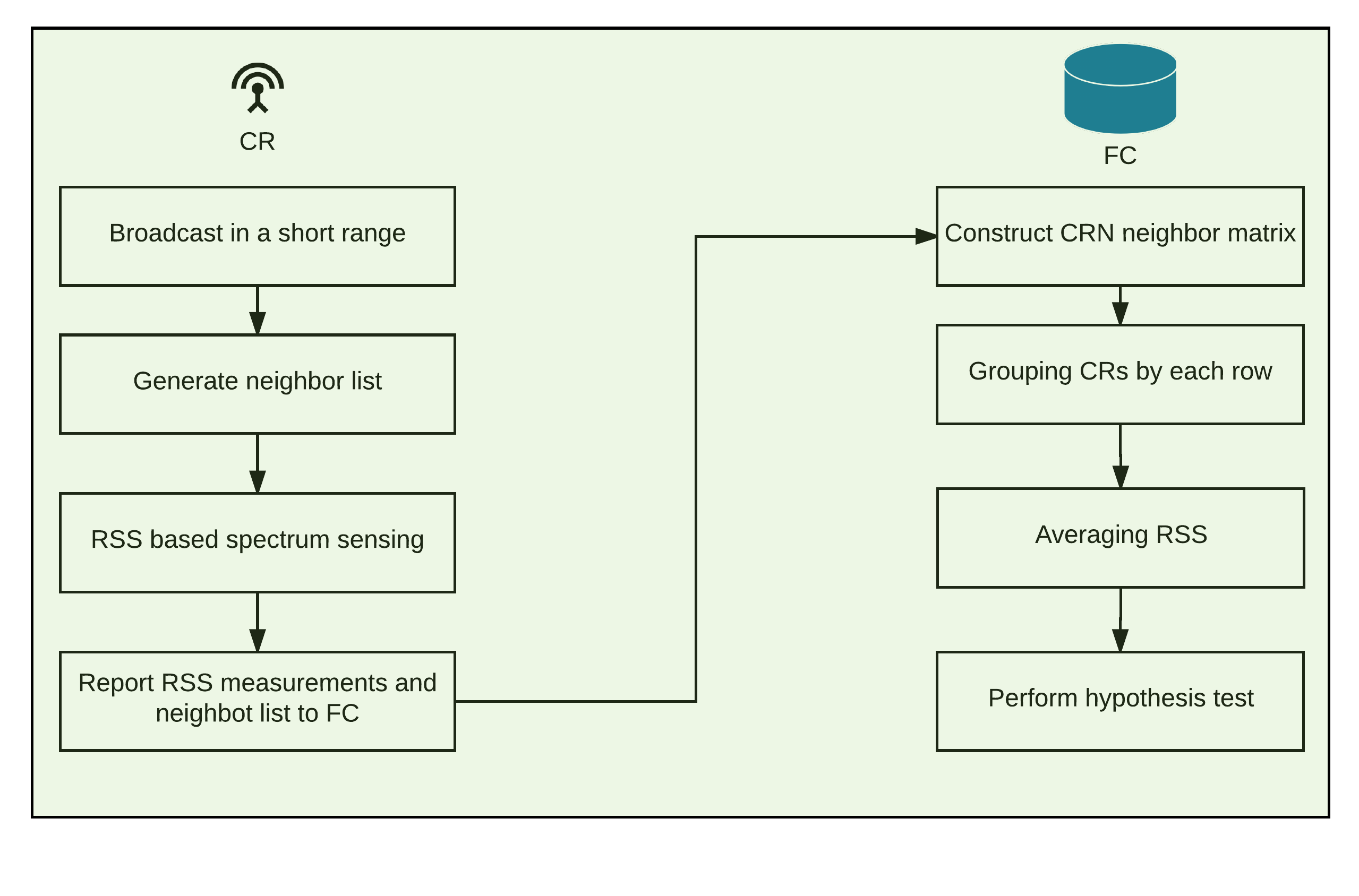}
	\caption{Process of CR grouping and PUE attack detection.}
	\label{fig:process}
\end{figure}

In comparison to clustering patterns in traditional wireless sensor networks (WSNs), CRs grouping does not meant to construct a hierarchical CRN structure. Instead, it is a logical grouping process that is completed by the FC. The grouping process is shown in Fig. \ref{fig:process}. Every CR will maintain a dynamic neighbor list by intermittently requesting in a short broadcasting range $r_{neighbor}$. In spectrum sensing stage, CRs will send their neighbor list along with the RSS measurements to the FC, which enables the FC create a $N \times N$ binary CR neighbor matrix $A_{neighbor}$ with each element be denoted as $a_{i,j}$. The FC will group RSS measurements by rows (for every $cr(i)$), shown in Fig. \ref{fig:group}. In each group, the averaged propagation loss is computed as $L_{i}^\ast=P_{t,pu}-mean({P_{r,k}|\forall a_{i,k}=1})$. Further, the distance between the PU and each group is estimated as $d_{i,pu}$, when it assumes all CRs in a group have approximately the same distance to the PU, because $d_{i,pu} \gg r_{neighbor}$.

\begin{figure}[t]
	\centering
		\includegraphics[width=0.7\textwidth]{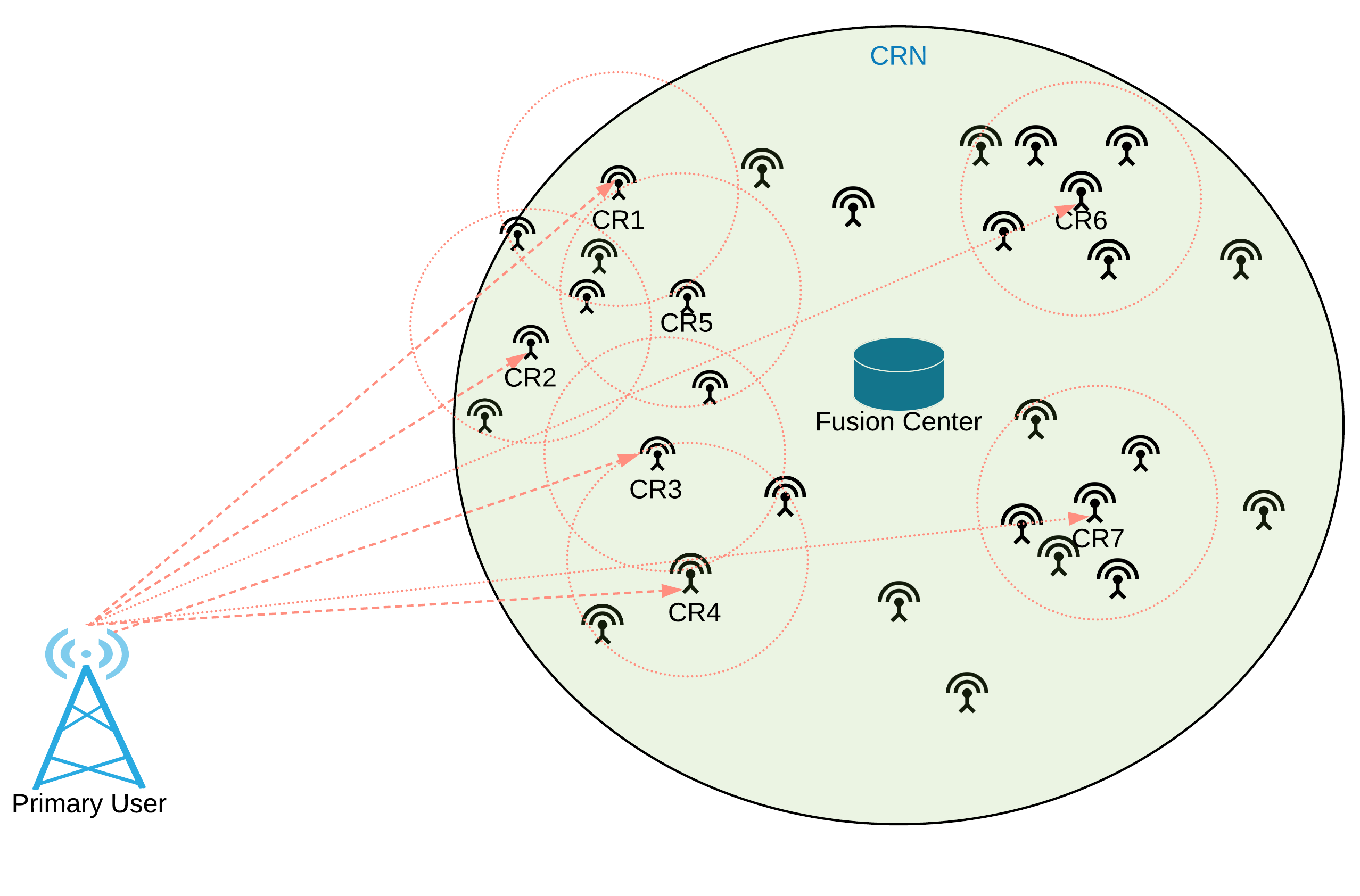}
	\caption{CRs grouping and RSS smoothing diagram.}
	\label{fig:group}
\end{figure}

\subsection{Hypothesis test of PUE attack detection}

In practical PUE attack detection, the hypothesis test defined in Section \ref{sec:perfect} is adopted. The propagation powers of the PU and attacker are denoted as $P_{t,pu} (dBm)$ and $P_{t,attacker} (dBm)$, respectively, where the propagation power difference, regarded as radiometric difference, is calculated as $F (dB)=P_{t,pu}-P_{t,attacker}$.

Refer to Eq. \ref{eq:model}, the distance between a transmitter and the $i$-th CR group is estimated as:

\begin{equation}
\hat{d}_{i} = 10^{(L_{i}^\ast-C-\varepsilon)/\Gamma}
\end{equation}

\noindent{where $\varepsilon$ is the ramaining error term. If the signal is transmitted by the PU, the estimated distance is the approximate distance between $i$-the CR and the PU $\hat{d}_{i,pu}$:}

\begin{equation}
\label{eq:Dpu}
\hat{d}_{i} = \hat{d}_{i,pu}
\end{equation}

If the signal is transmitted by the attacker, the path loss is computed as:

\begin{align}
L_{i}^\ast=&(P_{t,attacker} +F)- mean({P_{r,k}|\forall A_{i,k}=1})
\end{align}

Thus, the estimated distance is a scaled approximate distance between $i$-the CR and the attacker $\hat{d}_{i,attacker}$:
\begin{equation}
\label{eq:Dattacker}
\hat{d}_{i} = \hat{d}_{i,attacker}\cdot 10^{F/\Gamma}
\end{equation}

As assumed in section \ref{sec:perfect}, all CRs are randomly distributed in a circular area with radius of $r_{CRN}$. The transmitter-receiver distances satisfy:

\begin{eqnarray}
d_{i,pu}-d_{j,pu} \le 2 \cdot r_{CRN}, \forall i,j = 1,2,\cdots, N  \\
(d_{i,attacker}-d_{j,attacker})\cdot 10^{F/\Gamma} \le 2 \cdot r_{CRN}\cdot 10^{F/\Gamma}, \nonumber \\
\forall i,j = 1,2,\cdots, N
\end{eqnarray}

Refer to Eq. \ref{eq:Dpu} and Eq. \ref{eq:Dattacker}, the FC can apply the following hypothesis test:

\begin{itemize}
\item If $\hat{d}_{i} -\hat{d}_{j} \le T, \forall i,j = 1,2,\cdots, N$, the signal is from the PU ($\mathcal{H}_0$), or
\item If $\hat{d}_{i} -\hat{d}_{j} > T, \exists i,j = 1,2,\cdots, N$, the signal is from the attacker ($\mathcal{H}_1$)
\end{itemize}

Here $T$ is the threshold factor that affects the accuracy of the hypothesis test. The probability of false negative can be calculated as:

\begin{align}
\label{eq:fn}
P_{fn} =& \Pr \{ max(\hat{d}_{i,attacker} )-min(\hat{d}_{j,attacker})\le \frac{T}{10^{F/\Gamma} },\nonumber \\
 &\forall i,j = 1,2,\cdots, N\} \nonumber \\
&\left\{
\begin{array}{ll}
      \le&1,\text{ if } \frac{T}{10^{(F+\varepsilon^\prime)/\Gamma}} > r_{CRN}\\
      \le &(1- (\frac{r_{CRN}-\frac{T}{10^{(F+\varepsilon^\prime)/\Gamma}}}{r_{CRN}})^2)^N , \nonumber \\
& \text{ if }\frac{T}{10^{(F+\varepsilon^\prime)/\Gamma}} \le r_{CRN} \\
\end{array}
\right. \\
\end{align}

\noindent{where $\varepsilon^\prime$ is the error term. The interpretation to Eq. \ref{eq:fn} is similar to the one to Eq. \ref{eq:Prd}. It is noteworthy that the equality happens only when attacker is located at some particular locations. The probability of false positive can be calculated as:}

\begin{align}
\label{eq:fp}
P_{fp} &= \Pr \{ max(\hat{d}_{i,pu} )-min(\hat{d}_{j,pu})> T,\nonumber \\
 &\exists i,j = 1,2,\cdots, N\} \nonumber \\
&\left\{
\begin{array}{ll}
      =&0 , \text{ if }\frac{T}{10^{\varepsilon^\prime/\Gamma}} \ge 2 \cdot r_{CRN} \\
      < &1- (\frac{\alpha-\frac{T}{10^{\varepsilon^\prime/\Gamma}}}{\pi})^N, \nonumber \\
  & if\ \frac{T}{10^{\varepsilon^\prime/\Gamma}} < 2 \cdot r_{CRN}\\
\end{array}
\right. \\
\end{align}

\noindent{where $\cos \alpha=\frac{r_{CRN}-T/10^{\varepsilon^\prime/\Gamma}}{r_{CRN}}$. The Eq. \ref{eq:fp} can be explained as the complementary of the probability to the case that all CRs are located in the intersection area between a ring-shape area with width of $\frac{T}{10^{\varepsilon^\prime/\Gamma}}$ and the CRN distributed area. According to Eq. \ref{eq:fn} and Eq. \ref{eq:fp}, with larger value of $F$ and lower value false negative rate $P_{fn}$, better hypothesis threshold factor $T$ can be designed. With the larger number of CRs $N$, the lower false negative rate $P_{fn}$ can be achieved, but a higher false positive rate $P_{fp}$ may occur.}

\section{Real-world Emulation Trial}
\label{sec:experiment}
In this section, a deployment trail of our method in real-world PUE attack detection is presented. To perform spectrum sensing in CRN, we used Universal Software Radio Peripheral (USRP) N210 as the sensing nodes, one
of which acts as a smart PUE attacker. Due to the practical limitations, we are unbale to emulate PU activities. Thus, we regard one of the local digital television (DTV) station as the primary user. The PUE attacker impose malicious signal on another unused spectrum band. In order to conduct effective attacks, the smart attacker will mimic the DTV behavior: it will record the DTV signal from near spectrum band and broadcast the exact received signal data.

We implemented the experiment in our lab. The attacker (one USRP N210) is allocated to a fixed spot, and the sensing nodes (other USRP N210 Devices) are placed in 6 different places/rooms, shown in Fig. \ref{fig:floor}. Due to lack of empirical model parameters, we directly applied Hata propagation model for urban environment~\cite{wirelessCommunication}. The PU signal information is presented in Table \ref{tab:pu}, where $h_T$ is the transmitter height. Accordingly, we take of the value of receiver height $h_R$ as $10$m.

\begin{figure}[t]
	\centering
		\includegraphics[width=0.8\textwidth]{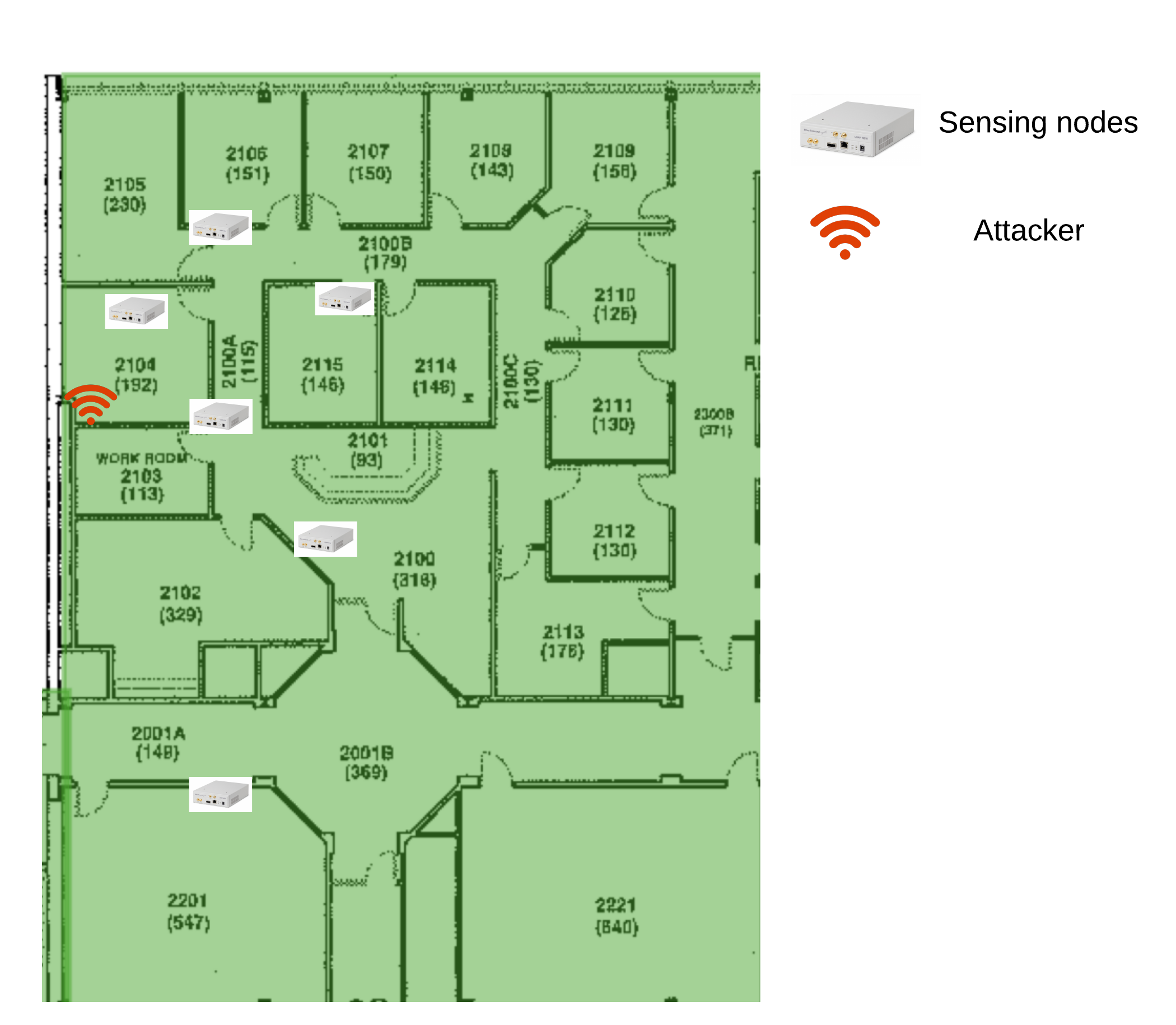}
	\caption{Experiment deployment.}
	\label{fig:floor}
\end{figure}

\begin{table}[ht]\centering
\caption{PU parameters}
\label{tab:pu}
    \begin{tabular}{p{30mm}|p{30mm}}
    \hline
         Frequency  & $590-596$ MHz \\
        	Power &  345 kW  \\
        $h_T$   &   $278$ m	\\
         \hline
    \end{tabular}
\end{table}

The result is shown in Fig \ref{fig:emulation}, which indicates an almost perfect detection. It is because the great discrepancy between PU transmission power and attacker transmission power (over 60 dB difference), despite the inaccurate propagation model parameters. The sensing nodes will receive a relatively high power of PUE attack signal if near to the attacker, but receive barely nothing if too far away from the attacker. In next section, we will present more detail discussions on detection performance regarding to model parameter errors and attacker transmission power.

\begin{figure}[t]
	\centering
		\includegraphics[width=0.6\textwidth]{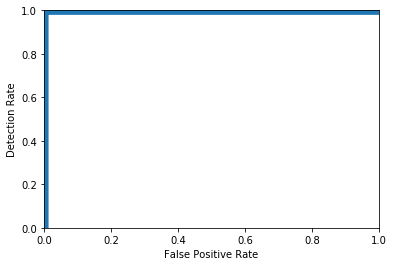}
	\caption{ROC of emulation.}
	\label{fig:emulation}
\end{figure}

\section{Numerical evaluation}
\label{sec:evaluate}

\subsection{Practical Model Evaluation}
Further, a numerical experiment with more comprehensive network topology, is designed to evaluate the proposed hypothesis test. The parameters in Eq. \ref{eq:model} is estimated from empirical study, which may not be the best estimation. The empirical model we used for distance estimation is:

\begin{equation}
\label{eq:em_model}
L=  C_{est}+\Gamma_{est} \cdot log_{10} d+ G
\end{equation}

While the best fit propagation model is:

\begin{align}
\label{eq:puReal}
L= &  (C_{est}-\varepsilon\_C)+(\Gamma_{est}-\varepsilon\_\Gamma) \cdot log_{10} d+ G \nonumber \\
	= & C_{est} + \Gamma_{est}\cdot (log_{10} d-\frac{\varepsilon\_C}{\Gamma}-\frac{\varepsilon\_\Gamma}{\Gamma}log_{10} d)
\end{align}

\noindent{where $\varepsilon\_C$ and $\varepsilon\_\Gamma$ are the empirical propagation model estimation errors ($C_{est}-C_{best}=\varepsilon\_C$ and $\Gamma_{est}-\Gamma_{best}=\varepsilon\_\Gamma$). Thus, the estimated distance is, if signal is transmitted by the PU:}

\begin{equation}
\hat{d}_{i}= (\hat{d}_{i,pu})^{(1-\varepsilon\_\Gamma/\Gamma)}\cdot 10^{-\varepsilon\_C/\Gamma}
\end{equation}

Similarly, the estimated distance calculated based on the attacker transmission signal is:

\begin{equation}
\hat{d}_{i}= (\hat{d}_{i,attacker})^{(1-\varepsilon\_\Gamma/\Gamma)}\cdot 10^{(-\varepsilon\_C/\Gamma+F/\Gamma)}
\end{equation}

Compared to Eq. \ref{eq:fn} and Eq. \ref{eq:fp}, the empirical propagation model estimation errors may increase both the false positive and false negative probabilities, due to the increasing uncertainty from the estimated distance.

\subsection{Numerical Test and Comparison}
The designed test scenario is in a $3000m\times3000m$ field. The PU and the FC are initially randomly located in the field. The PU is able to move. CRs and the attacker are randomly distributed in a circular area with radius $500m$. The best fitted propagation model parameters, $C_{best}$ and $\Gamma_{best}$, are designed by refering to the empirical Hata model \cite{wirelessCommunication}. The model parameter errors follow Gaussian distribution, defined as $\varepsilon\_C\sim(0,\sigma_{\varepsilon\_C}^2)$ and $\varepsilon\_\Gamma\sim(0,\sigma_{\varepsilon\_\Gamma}^2)$. The details are shown in Table \ref{tab:para}.

\begin{table}[ht]\centering
\caption{Parameter Setting}
\label{tab:para}
    \begin{tabular}{p{30mm}|p{30mm}}
    \hline
         Field  & $3000m\times3000m$ \\
        	$C_{best}$ &  111.76  \\
        $\Gamma_{best}$   &   31.8	\\
	$G$ &  $\sim(0,8^2)$  \\
	$\varepsilon\_C$   &   $\sim(0,\sigma_{\varepsilon\_C}^2)$ 	\\
	$\varepsilon\_\Gamma$   &   $\sim(0,\sigma_{\varepsilon\_\Gamma}^2)$ 	\\
	$P_{t,pu}$   &   $50 (dBw) = 100(W)$ 	\\
	PU Mobility &  	Yes				\\
         \hline
    \end{tabular}
\end{table}

We have compared the performance of our proposal with a back propagation neural network (BPNN) based approach \cite{Peng:2014aa}. It is a PUE attack detection scheme that does not need geographical information of the PU, which is similar to our work. However, it does require CRs' geographical information for both training and testing process. Although there are other PUE attacks detection methods, their strong assumptions make it inappropriate to compare them with our approach. In the evaluation test, we apply a three layer BPNN with three input nodes, four hidden nodes and two output nodes, as shown in Fig. \ref{fig:bpnn}.

\begin{figure}[t]
	\centering
		\includegraphics[width=0.55\textwidth]{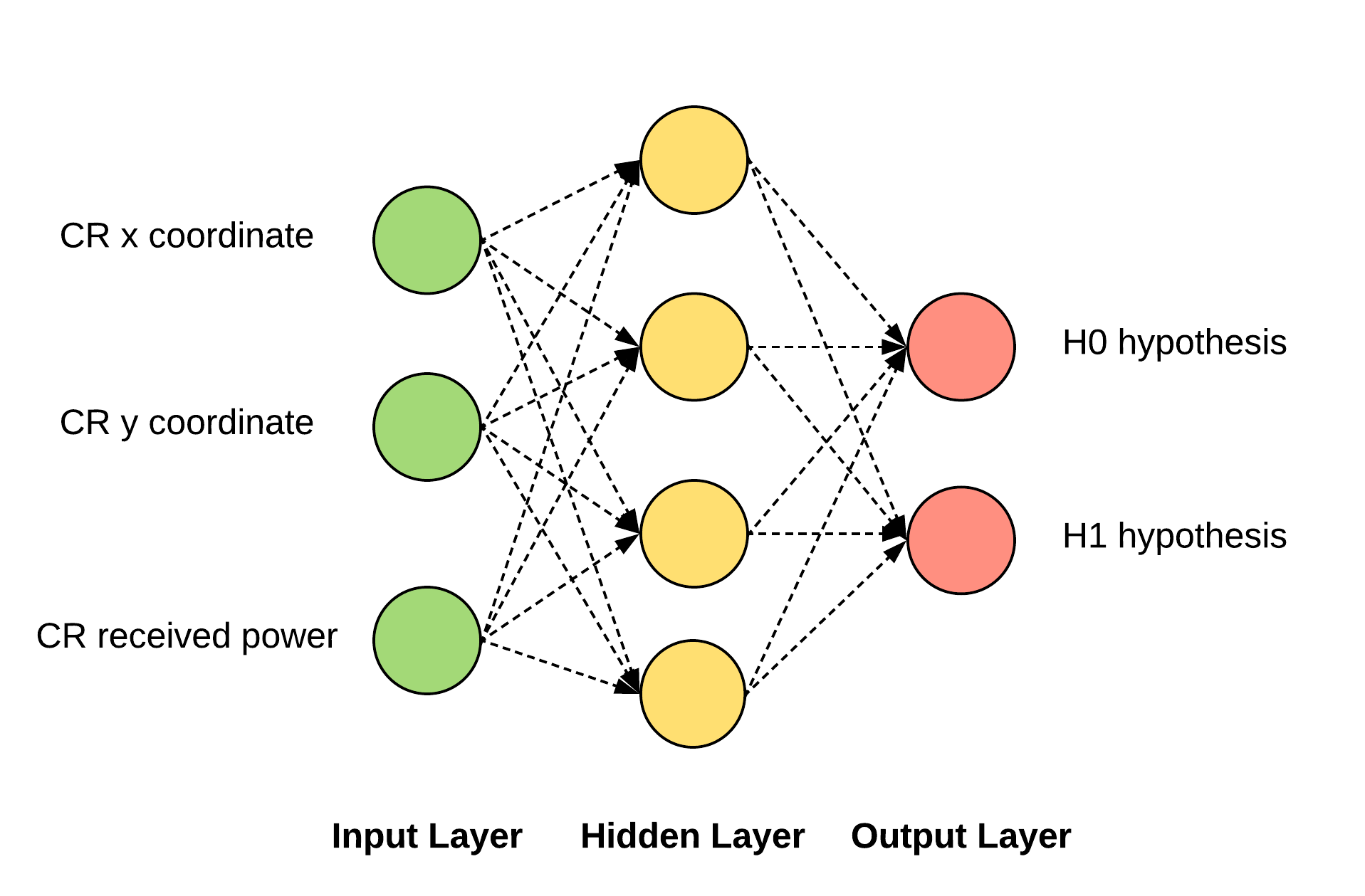}
	\caption{BPNN structure for PUE attack detection.}
	\label{fig:bpnn}
\end{figure}

\begin{figure}[t]
	\centering
		\includegraphics[width=0.55\textwidth]{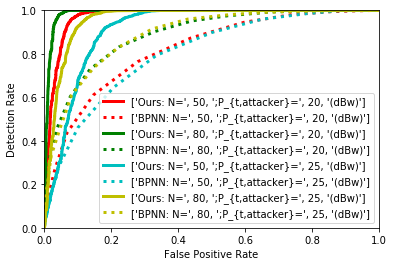}
	\caption{ROC of two approaches,, when $\sigma_{\varepsilon\_C}=0$ and $\sigma_{\varepsilon\_\Gamma}=0$, with different number of CRs ($N$) and different attacker propagation power $P_{t,attacker}$.}
	\label{fig:resultComp00}
\end{figure}

\begin{figure}[t]
	\centering
		\includegraphics[width=0.55\textwidth]{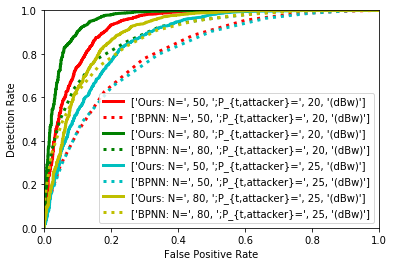}
	\caption{ROC of two approaches,, when $\sigma_{\varepsilon\_C}=3$ and $\sigma_{\varepsilon\_\Gamma}=1$, with different number of CRs ($N$) and different attacker propagation power $P_{t,attacker}$.}
	\label{fig:resultComp31}
\end{figure}

\begin{figure}[t]
	\centering
		\includegraphics[width=0.55\textwidth]{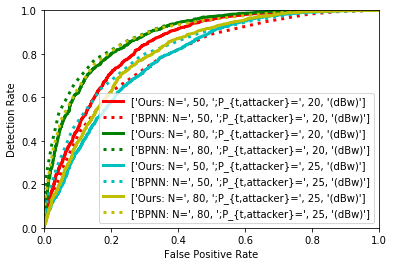}
	\caption{ROC of two approaches, when $\sigma_{\varepsilon\_C}=5$ and $\sigma_{\varepsilon\_\Gamma}=2$, with different number of CRs ($N$) and different attacker propagation power $P_{t,attacker}$.}
	\label{fig:resultComp52}
\end{figure}

\begin{figure}[t]
	\centering
		\includegraphics[width=0.6\textwidth]{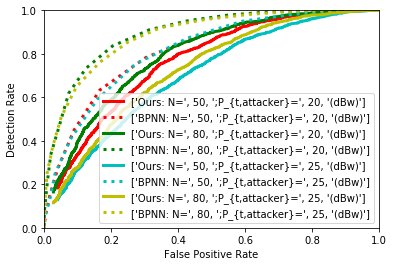}
	\caption{ROC of two approaches,, when $\sigma_{\varepsilon\_C}=10$ and $\sigma_{\varepsilon\_\Gamma}=3$, with different number of CRs ($N$) and different attacker propagation power $P_{t,attacker}$.}
	\label{fig:resultComp103}
\end{figure}

Figures \ref{fig:resultComp00}, \ref{fig:resultComp31}, \ref{fig:resultComp52}, and \ref{fig:resultComp103} present the comparison between our proposal and the BPNN approach using the receiver operating characteristics (ROC) curves corresponding to different number of CRs ($N$) and different propagation power differences ($F$) under several different parameter error propagation models.

The performance evaluation results in the figures show that both our proposed approach and BPNN approach for PUE attack detection have achieved better performance when there are larger number of CRs and larger propagation power difference between the PU and the attacker. When compared all result figures, however, it is shown that the BNPP approach is not sensitive to model parameter errors $\sigma_{\varepsilon\_C}$ and $\sigma_{\varepsilon\_\Gamma}$, while the performance of our approach greatly depends on the accuracy of model estimation. It is because the training data feeding to the neural network in BPNN approach is directly from real propagation environment, thus the testing process does not rely on the propagation model estimation. As shown in Figs. \ref{fig:resultComp00} and \ref{fig:resultComp31}, on the other hand, our approach achieves a superior performance when the propagation model is well estimated.

However, the comparison based only on performance does not provide a comprehensive vision. The BPNN is robust against the inaccuracy in propagation model estimation because it is essentially empirical and learns from historical data. Actually the BPNN detector does not work with the same inputs that are required by our proposed method.

In summary, our proposed detection approach possesses two major advantages over the BPNN detector. Firstly, the BPNN approach requires CRs' geographical information in both training and testing process, which may greatly increase the cost by equipping CRs with extra peripherals, such as GPS, while our approach does not rely on any prior geographical information. Secondly, in our approach, no training process, especially supervised training process, is required. In PUE attack detection, training signal at receiver sides with tag of the PU is not always available in practical. Therefore, our approach, compared to BPNN detector, is more feasible in a wide selection of scenarios.

\section{Conclusions}
\label{sec:conclusions}
In this work, we proposed a novel PUE attack detection approach leveraging the hard-to-mimic feature of high PU transmission power, compared to the attacker transmission capability. The detection model considered many constraints in real-world situations, such as mobile PUs, unknown geographical information of each party, and the geographical randomness of PUs and attackers as well as the CRN formation. Both theoretical analysis and experimental results have validated our proposal.


\section*{Acknowledgement}

Q. Dong, Y. Chen and X. Li are supported by the NSF via grant CNS-1443885. K. Zeng is partially supported by the NSF under grant No. CNS-1502584
and CNS-1464487.



%

\footnotesize

\bibliographystyle{splncs_srt}
\bibliography{CRN_DQ.bib}

\end{document}